\documentstyle[aps,prl,graphicx,floats]{revtex}
\newcommand{\beq}{\begin{equation}}
\newcommand{\eeq}{\end{equation}}

\begin{document}

\baselineskip=18pt

\begin{center}
{\Large\bf Structural transitions  in biomolecules - a numerical 
           comparison of two approaches for the study of
           phase transitions in small systems}

\vskip 1.1cm
{\bf Nelson A. Alves\footnote{E-mail: alves@quark.ffclrp.usp.br}}
\vskip 0.1cm
{\it Departamento de F\'{\i}sica e Matem\'atica, FFCLRP 
     Universidade de S\~ao Paulo. Av. Bandeirantes 3900. \\
     CEP 014040-901 \, Ribeir\~ao Preto, SP, Brazil}
\vskip 0.3cm
{\bf Ulrich H.E. Hansmann \footnote{E-mail: hansmann@mtu.edu} ~~
     ~and~~ Yong Peng\footnote{E-mail: ypeng@mtu.edu}}
\vskip 0.1cm
{\it Department of Physics, Michigan Technological University,
         Houghton, MI 49931-1291, USA}

\vskip 0.3cm
\today
\vskip 0.4cm
\end{center}
\begin{abstract}
We compare two recently proposed methods for the characterization of
phase transitions in small systems. The usefulness of these techniques
is evaluated for the case of structural transition in alanine-based
peptides.

 
\end{abstract}  
\vskip 0.1cm
{\it PACS-No.: 05.70.Fh, 05.50.+q, 64.60.-i, 87.15.He}



\nopagebreak


\section{INTRODUCTION} 
\noindent
When phase transitions are studied in statistical physics, the basic 
assumption is usually that the dimensions of the macroscopic system 
are very large  compared with that of the constituting elements. 
However, there are  many phenomena in finite systems 
which  resemble phase transitions, for instance, 
in the physics of clusters of atoms \cite{Berry}.
The main 
question is  how the observed effects in small systems can be related
to true phase transitions in macroscopic (or infinite) systems. 
Attempts in this direction include the exploration 
of the density of complex zeros of the 
canonical partition function for finite and small systems 
by Borrmann et. al. \cite{BMH}
and the linear behavior for this limiting density
\cite{GR1,GR2}.
The approach suggested by Janke and Kenna (JK) presents also a   
new scaling relation to identify the order and  strength of a 
transition  from the behavior of small systems \cite{JK}.

In this paper we try to evaluate the usefulness and validity 
of the Borrmann and the JK approaches for the investigation 
of structural transitions in biomolecules, another important
example for ``phase transitions'' in small systems. For this purpose 
we consider the helix-coil transition in polyalanine and test 
whether the two approaches allow a characterization of 
that transition. In addition, we also investigate a second molecule
Ala$_{10}$-Gly$_5$-Ala$_{10}$ in the context of Borrmann approach.

\section{ANALYSIS OF PARTITION FUNCTION ZEROS IN SMALL SYSTEMS}
\noindent
 In the canonical ensemble a system is completely described 
 by its partition function
\beq
  Z(\beta) = \sum_E n(E) {\rm exp} (-\beta E) \, .
\eeq
Introducing the variable $u = \exp (-k\beta)$ with conveniently
defined constant $k$ allows to write the partition function 
for discrete energy models as a polynominal:
\beq
  Z(u) = \sum_E n(E) u^{E}~.                            \label{z2}
\eeq
We label the complex zeros in order of increasing imaginary part as
$u_j = {\rm exp}(-k \beta_j), j=1,2, ...$.
 
As established by Yang and Lee \cite{YL}
and later by Fisher \cite{Fisher}, the statistical theory
of phase transitions can be described by the distribution of
complex partition function zeros.
In the case of a temperature driven phase transition, we expect the zeros
(or at least the ones close to the real axis) condense for 
large enough system size $L$ on a single line 
\beq 
u_j = u_c + r_j\,{\rm exp} (i \varphi)~.                   \label{a1}
\eeq
As the system size $L$ increases, these zeros will move towards
the positive real $u$-axis  and for large $L$ the corresponding 
value is $\beta_c$, the inverse of the physical critical temperature $T_c$.
Crucial information on phase transitions can be obtained from 
the way in which the first zero approaches the real $u$-axis.
However, such an analysis depends on the extrapolation
towards the infinite large system and does not allow  characterization 
of the situation in small systems. 

One possible extension of the above ideas to ``phase transitions'' in
small systems is the classification scheme by  Borrmann {\it et al.}
\cite{BMH}.
This classification scheme computes density of zeros taking into 
account numerical estimates of the first four complex zeros. 
Now, writing the
complex zeros  as $u_k = {\rm Re}(u_k) + i \, \tau_k$, 
where $\tau_k$ stands for ${\rm Im}(u_k)$,
the assumed
distribution of zeros on a straight line allows to define 
two parameters $\alpha_u$ and $\gamma_u$:
\beq
 \alpha_u = \frac{ {\rm ln}\,\phi(\tau_3) - {\rm ln}\,\phi(\tau_2)}
               { {\rm ln}\,\tau_3 - {\rm ln}\,\tau_2} \, ,    
\quad {\rm where} \quad
\phi(\tau_k) = \frac{1}{2} \left(\frac{1}{|{u}_k - {u}_{k-1}|}
         + \frac{1}{|{u}_{k+1} - {u}_{k}|}\right) \, ,        \label{phi}
\eeq
with $k$ labelling the first zeros, and 
\beq
\gamma_u = ({\rm Re}(u_2)-{\rm Re}(u_1))/(\tau_2 - \tau_1) \,.  \label{gamma}
\eeq
 Note, that our notation differs from that in Ref.~\cite{BMH} in that
 we define the discrete line density $\phi$ in function of the 
$u$-zeros instead of the inverse temperature $\beta$.   
Following the classification scheme by Grossmann 
and Rosenhauer \cite{GR1,GR2},
phase transitions can now be classified according to the values of
these two parameters:
for $\alpha_u = 0$ and $\gamma_u = 0$ one has a phase transition of
first order, it is of second order if $0 < \alpha_u < 1$ and 
arbitrary $\gamma_u$, and for $\alpha_u > 1$ 
and arbitrary $\gamma_u$ one has a higher order transition. 
 Of course, as expected for a phase transition, the imaginary part
of the leading zero $(\tau_1)$ should move towards the real axis
to obtain in the thermodynamic limit a real temperature.
However, for small systems with finite $\tau_1$, we may have
$\alpha < 0$ \cite{BMH}, which translates the condition of 
first order transition to  $\alpha_u  \leq 0$ and $\gamma_u =0$. 
This classification of phase transitions
has been tested for the finite Bose-Einstein condensates in a harmonic trap,
small magnetic clusters and nuclear multifragmentation \cite{BMH}.

Another extension of partition function zero analysis to small systems
is the approach by Janke and Kenna (JK) \cite{JK}, which uses that  
the theoretical average cumulative density of zeros \cite{JK}
\beq
  G_L(r_j) = \frac{2j-1}{2L^d}                           \label{a2}  
\eeq
can be written in the thermodynamic limit and
for a first order transition as
\beq
G_{\infty}(r) = g_{\infty}(0)r + a r^{w+1} + \cdots \, .  \label{a3}
\eeq
Here,  the slope at the origin is related to the latent heat,
$\Delta e \propto g_{\infty}(0)$.
  Equations~(\ref{a1}) and (\ref{a3}) imply that the 
distance $r_j$ of a zero 
from its critical point can be written for  
large enough lattice sizes  as ${\rm Im}\,u_j(L)$ since
${\rm Re}\,u_j(L) \sim u_c$. Hence, in this limit 
Eq.~(\ref{a2}) and (\ref{a3}) lead   to the following 
scaling relation for the cumulative density of zeros
as an equation in $j$ and $L$,
\beq
\frac{2j -1 }{2L ^d} = a_1 ({\rm Im}\, u_j(L))^{a_2} + a_3 \,.\label{a5}
\eeq
A necessary condition for the existence of a phase transition 
is that $a_3$ is compatible with zero, else it would indicate
that the system is in a well-defined phase. The values of
the constants $a_1$ and $a_2$ then  characterize the phase transition. 
For instance, for first order transitions
should the constant $a_2$  take values $a_2 \sim 1$ for small $r$,
 and in this case the slope of this equation is
related to the latent heat through the relation \cite{JK}
\beq
  \Delta e = k\, u_c\, 2 \pi a_1  \,,                 \label{a7}
\eeq
with $u_c = {\rm exp}(-k \beta_c)$. 
On the other hand, a value of $a_2$ larger than $1$
indicates a second order transition whose specific heat
exponent is given by $\alpha = 2 -a_2$.

The above approach was originally developed and tested for
systems with well defined first order phase transitions such as the  
2D 10-state Potts and 3D 3-state Potts model.
The obtained results agree with previous work from numerical 
simulations and partition function zeros analysis of systems up to $L=64$ 
\cite{Thesis,AlvesNB20}  ($L=36$ for the 3D case \cite{AlvesB43}).
However, some difficulties in the identification
of the order of the transition appear when applied to 
a model with a very weak first order transition \cite{AH-potts}.

 Here we explore further these two approaches for a non-trivial
example of structural transitions in biomolecules, namely the helix-coil
transition in polyalanine.  The characteristics of this so-called 
helix-coil transition have been studied extensively \cite{Poland}. 
For instance, evidence was presented \cite{HO98c,AH99b,KHC99c} that
 polyalanine exhibits a phase transition between the ordered
helical state and the disordered random-coil state when interactions
between all atoms in the molecule are taken into account. These previous
results, obtained independently by different methods, are used here 
as benchmarks for the test of the usefulness
and to point out the difficulties in applying the 
classification scheme of Borrmann and Janke/Kenna for research of
structural transitions in biomolecules.

Our investigation of the helix-coil transition for polyalanine is
based on a detailed, all-atom representation of that homopolymer.
Since one can avoid the complications of electrostatic and
hydrogen-bond interactions of side chains with the solvent for alanine
(a nonpolar amino acid), explicit solvent molecules were neglected.
The interaction between the atoms  was
described by a standard force field, ECEPP/2,\cite{EC}. 
 Chains of up to $N=30$ monomers were
considered, and our results rely on multicanonical simulations \cite{MU}
of $N_{sw}$ Monte Carlo sweeps starting from a random
initial conformation, i.e. without introducing any bias.
Our statistics consists of
$N_{sw}$=400,000, 500,000, 1,000,000, and 3,000,000 sweeps
for $N=10$, 15, 20, and 30, respectively. 
Measurements were taken every fourth Monte Carlo sweep.
Additional  40,000 sweeps ($N=10$)
to 500,000 sweeps ($N=30$) were needed for the  weight factor 
calculations by the iterative  procedure described first in
Ref.~\cite{MU}.  In contrast to our first calculation of complex zeros
presented in Ref.~\cite{AH99b}, where we divided the energy range into
intervals of lengths 0.5 kcal/mol in order to make Eq.~(\ref{z2})
a polynomial in the variable $u=e^{-\beta/2}$, we avoided  any 
approximation scheme in the present work. 
  This because the above 
approximation works very well for the first zero, but not
for the next ones. Since we  need high precision estimates
also for the next zeros we applied  the scan method
(see Ref. \cite{AlvesIJMPC} and references therein).

In Table 1  we present our first four partition function zeros,
although the fourth one is less reliable due to the presence of 
fake zeros \cite{AlvesNPB92}.
 Using these zeros we first calculated 
the parameters $\alpha_u$ and $\gamma_u$ which characterize
in the Borrmann approach phase transitions in small systems. As one
can see in Table 2 the so obtained values for polyalanine are characterized
by large fluctuations. It seems that the median of the $\alpha_u$ values 
is $\alpha_u = 0$ which would indicate a first order transition. However,
our data are not good enough to draw such a conclusion.
 Our error estimates were obtained by means of the Bootstrap method
\cite{Efron,AH-potts}.
  For this reason, we tried instead  the JK scaling relations.
Table 3 lists the parameter $a_3(N)$ of Eq.~(\ref{a5}).  
Here, the average cumulative density of zeros is replaced by
\beq
  G_N(r_j) = \frac{2j-1}{2N}  \, ,                      \label{poly1}  
\eeq
where we have translated the linear length $L$ 
as $N^{1/d}$ ~\cite{AH99b}. Therefore all finite size
scaling relations can be written in terms of the number of monomers $N$.
 In Fig. 1 we show the cumulative distribution of zeros in a log-log
plot for chain lengths $N=10, 15, 20 ~{\rm and}~30$.

The values $a_3$ are compatible with zero for chains of all length
indicating that we have indeed a
phase transition. In order to evaluate the kind of transition we also
calculate the parameters $a_1(N)$ and $a_2(N)$ which we also 
summarize in Table 3. Our analysis procedure differs from JK in that
we are calculating $a_2$ for a fixed $N$before performing a finite-size scaling
analysis of that quantity, i.e. we account for the dependence of the average 
cumulative density on the system size.  The parameter $a_2(N)$  
decreases with system size  and a log-log plot of this quantity as a
function of chain length, as shown in Fig. 2,
 suggests a scaling relation
\beq
a_2(N) = a_2 + b N^{-c} \, .                              \label{poly2}  
\eeq
A numerical fit of our data to this function leads to 
a value of $a_2 =1.31(4)$ with goodness of fit $Q=0.95$. 
Using $\alpha = 2 - a_2$ we find $\alpha =0.70(4)$
which is barely compatible with our previous value of $\alpha =0.86(10)$ in 
Ref.~\cite{AH99b}, obtained from the maximum of specific heat. 
A fit of all four chain lengths can also not exclude a value $a_2 =1$
(i.e. a first order phase transition)
since we can find acceptable fits with $Q > 0.55$ in the 
range $ 0.92 < a_2 < 1.44 $. Especially,
a close examination of Fig.~1 and Fig.~2 shows that the $N=30$ 
data point exhibits a 
considerable deviation from the trend suggested by the smaller chain lengths.
Since the $N=30$ data are the least reliable, we also evaluated 
Eq.~(\ref{poly2}) omitting the $N=30$ chain. This leads to a value of
$a_2=1.16(1)$ and a critical exponent $\alpha=0.84(1)$ which is now 
compatible with our previous result $\alpha = 0.86(10)$. 

While the JK approach is able to reproduce results for polyalanine obtained
in previous work \cite{AH99b}, it does not allow
to establish the order of the helix-coil transition from simulation
of small chains. Our results for the parameter $a_2$ seem to favor a
second order transition, but are hampered by large errors (such that a
first order transition cannot be excluded) and disputed in 
Ref.~\cite{KHC99c} were indications for a finite latent heat were found.

Borrmann's approach led to even less decisive results for polyalanine
since the fluctuation  were too large to draw firm conclusions.
However, the advantage of this approach is that it does not require
a finite-size scaling.
We have therefore used this method to investigate
a second molecule, the slightly more complicated 
 {\it Ala$_{10}$-Gly$_5$-Ala$_{10}$}. A detailed study of this
molecule will be published elsewhere \cite{AGA}.

We describe the interactions between
the atoms again by the  ECEPP/2 \cite{EC} force field (as
implemented in the program package SMMP \cite{SMMP}).
Our results rely on multicanonical simulations \cite{MU}
of $4,000,000$ Monte Carlo sweeps starting from a random
initial conformation, i.e. without introducing any bias.
Measurements were taken every tenth Monte Carlo sweep.
Additional  500,000 sweeps were needed for the  calculations
of the weight factor.  

Analyzing the partition function zeros for this peptide, we find 
a ``phase transition'' at {\it two} temperatures, each 
being characterized by a line of complex zeros.  The corresponding first four
zeros for each characteristic line are listed in Table 4.  Both temperatures
also correspond with peaks in the specific heat which is displayed
in Fig.~3 as a function of temperature. In Ref.~\cite{AGA} it is 
shown that the higher temperature $T_1=480$ K  marks the helix-coil
transition, i.e. the temperature where secondary structure elements
($\alpha$-helices) are formed. At the lower temperature $T_2=260$ K,
the peptide then folds into its native structure, a U-turn-like bundle
of two (antiparallel) $alpha$-helices connected by a turn. Here, we
use the Borrmann approach to study these two transitions in more detail.
Using the zeros from Table 4 we  calculated 
the parameters $\alpha_u$ and $\gamma_u$ which characterize
in the Borrmann approach phase transitions in small systems. 
For the first transition, at $T=480$ K,
which corresponds to the above studied helix-coil transition  in polyalanine,
we find $\alpha_u=-1.46$ and $\gamma_u=0.49$. While the negative 
value of $\alpha_u$ indicates a first order transition, our results
do again not allow us to draw a firm conclusion on the nature of the
transition since  the values of the two parameters
$\alpha_u$ and $\gamma_u$ are subject to large
fluctuations when we vary the number of sweeps 
considered in the calculation of the zeros. 
This is not surprising since we were also
not able to establish clearly the order of the helix-coil transition in
in the previous example of polyalanine.  However, our data are more 
decisive in the
case of the second transition, at $T=260$ K, which marks the 
compactification and folding of the peptide. Here we find $\alpha_u=0.38$
and $\gamma_u=-1.07$. These values indicate  a second-order transition 
which is consistent with what one would expect for a transition between 
extended
and compact structures. We hope that by doubling the number of Monte Carlo
sweeps we will be able to verify this results and  also obtain reliable
values for the Borrmann parameter characterizing the helix-coil transition 
at $T=480$ K. These simulations are now under way.

\section{Conclusion}
Let us summarize our results. 
We have evaluated two recently proposed schemes for characterizing
phase transitions in small systems. Simulating  polyalanine molecules of
chain lengths up  $N=30$ residues by multicanonical Monte Carlo, we
calculated the partition function zeros of these molecules. Analyzing
these zeros by the  JK approach we were able to reproduce for polyalanine 
results obtained in previous work \cite{AH99b} while the Borrmann approach 
led here to inconclusive results. However, the later method seems 
more appropriated to study structural transitions in proteins since it 
does not require a finite size scaling. When applied to 
 {\it Ala$_{10}$-Gly$_5$-Ala$_{10}$}, the Borrmann approach allowed us 
to characterize
structural transitions in this molecule. However, our results are limited again
by the precision with which one can calculate  the first four partition function
zeros. While these numerical limitations restrict the applicability of
the Borrmann and the JK approaches, our results demonstrate that both methods
can be  useful tools for the study of ``phase transitions'' in biomolecules.

\ \\
\noindent
{\bf  Acknowledgements}\\
 U.H.E. Hansmann gratefully acknowledges support by research grant 
of the National Science Foundation (CHE-9981874), and 
N.A. Alves support by CNPq (Brazil).



\newpage

{\huge Figure Captions:} \\
\begin{description}
\item[Figure 1.] Distribution of zeros for the polyalanine model.
\item[Figure 2.] The parameter $a_2(N)$ for polyalanine molecules of 
                 length $N$ in a log-log plot.
\item[Figure 3.] The  specific heat $C(T)$ as a function of 
                 temperature for Ala$_{10}$-Gly$_5$-Ala$_{10}$.
\end{description}
\ \\
{\huge Table Captions:}\\
\begin{description}
\item[Table 1.] Partition function zeros for polyalanine.
\item[Table 2.] Estimates for the parameters
          $\alpha_u$ and $\gamma_u$ for polyalanine.
\item[Table 3.]  Estimates for the JK parameters 
         for polyalanine.
\item[Table 4.] Partition function zeros corresponding to the two
         transitions observed for 
         Ala$_{10}$-Gly$_5$-Ala$_{10}$.
\end{description}

\newpage
\cleardoublepage


\begin{table}[ht]
\renewcommand{\tablename}{Table}
\caption{\baselineskip=0.8cm Partition function zeros for polyalanine.}
\begin{center}
\begin{tabular}{lcccccccc}\\
\\[-0.3cm]
$~N$~~&Re$(u_1)$ &Im$(u_1)$    & Re$(u_2)$ & Im$(u_2)$ &Re$(u_3)$ 
      &Im$(u_3)$ &Re$(u_4)$    & Im$(u_4)$  \\
\\[-0.35cm]
\hline
\\[-0.3cm]
~10 &0.30530(12) &0.07720(14)  &0.2823(13) &0.13820(61)&0.2459(72)&0.1851(63)
    &0.172(11)   &0.2200(71)    \\
~15 &0.356863(61)&~0.053346(39)&0.34167(60)&0.10440(59)&0.3331(48)&0.1454(28)
    &~0.3067(81) &0.1689(32)    \\
~20 &0.374016(41)&~0.042331(45)&0.36161(27)&0.08109(24)&0.3569(27)&0.1154(13)
    &~0.3336(56) &0.1470(27)    \\
~30 &0.378189(19)&~0.027167(32)&~0.37399(14)&~0.05420(27)&0.3693(11)
    &0.0804(13)  &~0.35854(63) &0.1022(43)    \\
\end{tabular}
\end{center}
\end{table}


\begin{table}[ht]
\renewcommand{\tablename}{Table}
\caption{\baselineskip=0.8cm  Estimates  for the parameters
          $\alpha_u$ and $\gamma_u$ for polyalanine.}
\begin{center}
\begin{tabular}{cll}\\
\\[-0.3cm]
$~N$  &~~~$\alpha_u$ & ~~~$\gamma_u$  \\
\\[-0.35cm]
\hline
\\[-0.3cm]
~10   &-0.36(17) & -0.365(17) \\
~15   & 0.41(19) & -0.291(11) \\
~20   & 0.06(14) & -0.3229(78)\\
~30   & 0.19(14) & -0.1568(58) \\
\end{tabular}
\end{center}
\end{table}

\begin{table}[ht]
\renewcommand{\tablename}{Table}
\caption{\baselineskip=0.8cm  Estimates for the JK parameters  
         for polyalanine.}
\begin{center}
\begin{tabular}{lccc}\\
\\[-0.3cm]
$~N$ & $a_1$   &  $a_2$   &  $a_3$ \\
\\[-0.35cm]
\hline
\\[-0.3cm]
~10 & 6.17(60) &1.862(46) &  0.01(14)  \\
~15 & 4.37(19) &1.664(16) &  0.014(69) \\
~20 & 3.62(26) &1.558(24) &~-0.014(98) \\
~30 & 3.54(31) &1.473(30) &~-0.007(61) \\
\end{tabular}
\end{center}
\end{table}

\begin{table}[ht]
\renewcommand{\tablename}{Table}
\caption{\baselineskip=0.8cm Partition function zeros corresponding to 
the two transitions observed for Ala$_{10}$-Gly$_5$-Ala$_{10}$.}
\begin{center}
\begin{tabular}{cccccccc}\\
\\[-0.3cm]
 Re$(u_1)$     & Im$(u_1)$     & Re$(u_2)$     & Im$(u_2)$ &
 Re$(u_3)$     & Im$(u_3)$     & Re$(u_4)$     & Im$(u_4)$        \\
\\[-0.35cm]
\hline
\\[-0.3cm]
 0.349620 & 0.047937 & 0.361590 & 0.072467 &
 0.320531 & 0.097282 & 0.370090 & 0.132103 \\
 0.144820 & 0.039763 & 0.104362 & 0.077531 &
 0.074312 & 0.100345 & 0.033091 & 0.116098 \\
\end{tabular}
\end{center}
\label{table 4}
\end{table}


\newpage
\cleardoublepage

\begin{figure}[t]
\begin{center}
\begin{minipage}[t]{0.95\textwidth}
\centering
\includegraphics[angle=-90,width=0.72\textwidth]{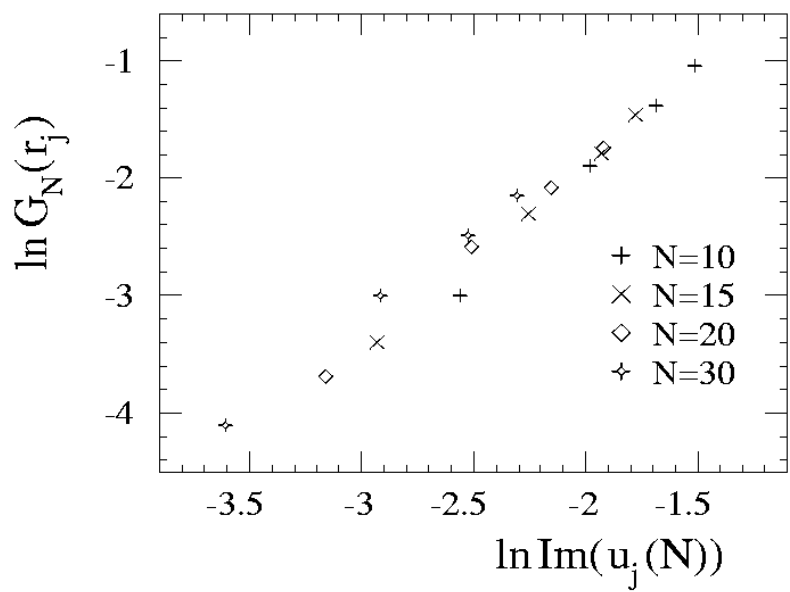}
\renewcommand{\figurename}{Fig.}
\caption{Distribution of zeros for the polyalanine model.}
\label{fig 1}
\end{minipage}
\end{center}
\end{figure}

\begin{figure}[!ht]
\begin{center}
\begin{minipage}[t]{0.95\textwidth}
\centering
\includegraphics[angle=-90,width=0.72\textwidth]{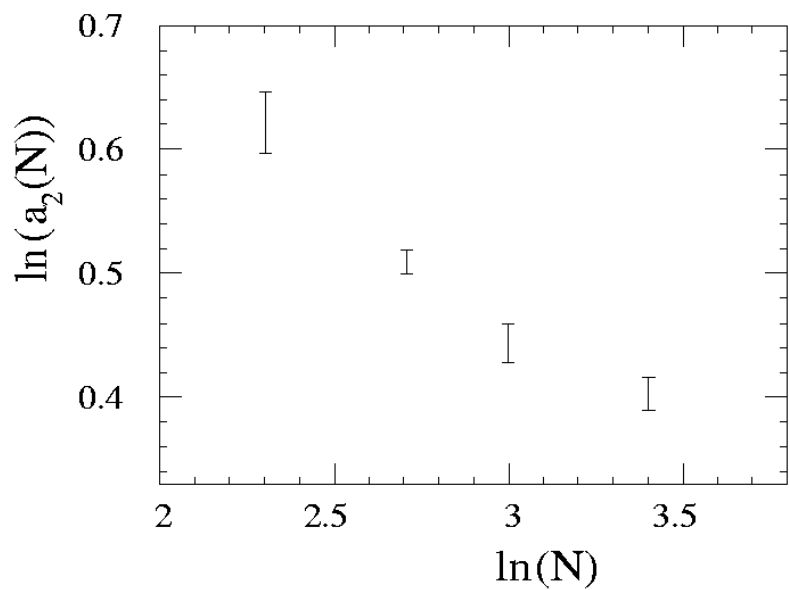}
\renewcommand{\figurename}{Fig.}
\caption{The parameter $a_2(N)$ for polyalanine molecules of length $N$
          in a log-log plot.}
\label{fig. 2}
\end{minipage}
\end{center}
\end{figure}

\begin{figure}[!ht]
\begin{center}
\begin{minipage}[t]{0.95\textwidth}
\centering
\includegraphics[angle=-90,width=0.72\textwidth]{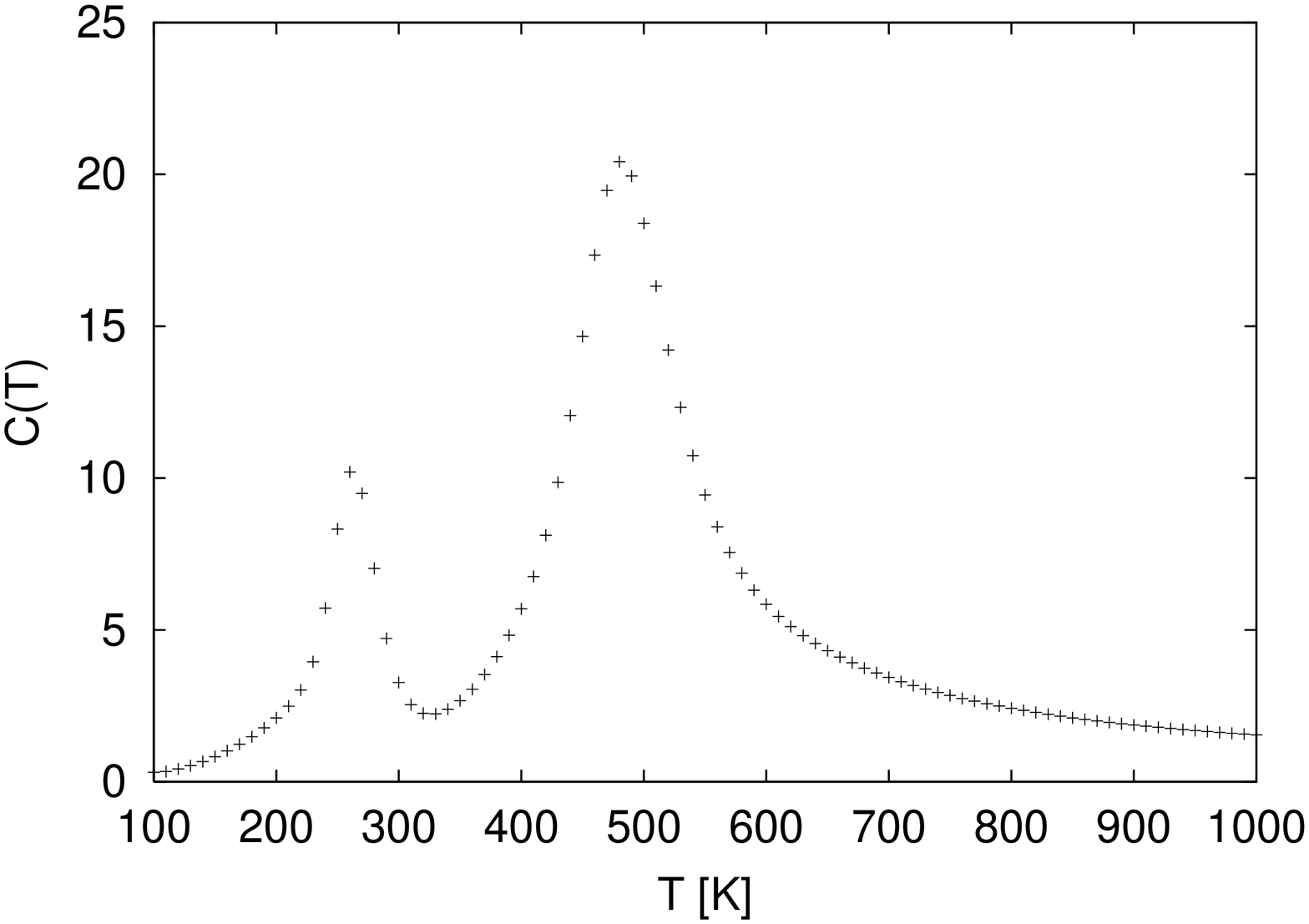}
\renewcommand{\figurename}{Fig.}
\caption{The  specific heat $C(T)$ as a function of temperature for
         Ala$_{10}$-Gly$_5$-Ala$_{10}$.}
\label{fig. 3}
\end{minipage}
\end{center}
\end{figure}


\begin{thebibliography}{99}
\bibitem{Berry} Berry, R.S.; Smirnov, B.M. Phase Stability of Solid Clusters. 
                    J. Chem. Phys. 2000, 113, 728-737.
                Berry, R.S.; Smirnov, B.M.
                  Structural Phase Transition in a Large Cluster. 
                  J. Exp. Theor. Phys. 2000, 90, 491-498.
                Proykova, A.; Radev, R.; Li, F.-Y.; Berry, R.S.
                   Structural Transitions in Small Molecular Clusters. 
                   J. Chem. Phys. 1999, 110, 3887-3896.
                Nayak, S.K.; Jena, P.; Ball, K.D.; Berry, R.S.
                   Dynamics and Instabilities Near the Glass Transition:
                   From Clusters to Crystals. 
                   J. Chem. Phys. 1998, 108, 234-239.
                Proykova, A.; Berry, R.S.
                   Analogues in Clusters of Second-Order Transitions ? 
                    Z. Phys. D. 1997, 40, 215-220.
                Wales, D.J.; Berry, R.S.
                   Coexistence in Finite Systems.
                   Phys. Rev. Lett. 1994, 73, 2875-2878.
                Kunz, R.E.; Berry, R.S.
                   Coexistence of Multiple Phases in Finite Systems.
                   Phys. Rev. Lett. 1993, 71, 3987-3990.
\bibitem{BMH} Borrmann, P.; M\"ulken, O.; Harting, J.
               Classification of Phase Transitions in Small Systems. 
               Phys. Rev. Lett. 2000, 84, 3511-3514.
              M\"ulken, O.; Borrmann, P.; Harting, J.; Stamerjohanns, H.
               Classification of Phase Transitions of Finite 
               Bose-Einstein Condensates in Power-Law traps by Fisher Zeros. 
               Phys. Rev. A. 2001, 64, 013611-1-013611-6.
              M\"ulken, O.; Borrmann, P.
                Classification of the Nuclear Multifragmentation 
                Phase Transition. 
               Phys. Rev. C. 2001, 63, 024306-1-024306-4.
\bibitem{GR1} Grossmann, S.; Rosenhauer, W.
              Temperature Dependence Near Phase Transitions in Classical
              and Quant. Mech. Canonical Statistics.
              Z. Physik. 1967, 207, 138-152.
\bibitem{GR2} Grossmann, S.; Rosenhauer, W.
              Phase Transitions and the Distribution of Temperature Zeros
              of the Partition Function.
              Z. Physik. 1969, 218, 437-448. 
              Grossmann, S.; Rosenhauer, W.
              Phase Transitions and the Distribution of Temperature Zeros
              of the Partition Function.
              Z. Physik. 1969, 218, 449-459. 
\bibitem{JK}  Janke, W.; Kenna, R. 
              The Strength of First and Second Order Phase Transitions  
              from Partition Function Zeroes.
               J. Stat. Phys. 2001, 102,  1211-1227.
\bibitem{YL} Yang, C.N.; Lee, T.D.
             Statistical Theory of Equations of State and Phase 
             Transitions. I. Theory of Condensation.
             Phys. Rev. 1952, 87, 404-409.
             Lee, T.D.; Yang, C.N.
              Statistical Theory of Equations of State and Phase 
              Transitions. II. Lattice Gas and Ising Model.
              Phys. Rev. 1952, 87, 410-419.
\bibitem{Fisher} Fisher, M.E. Lectures in Theoretical Physics.
             University of Colorado Press, Boulder. 1965; Vol. 7c, p 1.
\bibitem{Thesis} Villanova, R. Ph.D. Thesis (1991). 
                 Florida State University (unpublished).  
\bibitem{AlvesNB20} Villanova, R.; Alves, N.A.; Berg, B.A.
              Density of States and Finite Size Scaling Investigations.
              Nucl. Phys. B (Proc. Suppl.). 1991, 20, 665-668.
\bibitem{AlvesB43} Alves, N.A.; Berg, B.A.; Villanova, R.
             Potts Models: Density of States and Mass Gap from 
             Monte Carlo Calculations.
             Phys. Rev. B. 1991, 43, 5846-5856.
\bibitem{AH-potts} Alves, N.A.; Ferrite, J.P.N.; Hansmann, U.H.E.
             Numerical Comparison of Two Approaches for the
             study of Phase Transitions in Small Systems.       
             Phys. Rev. E (2002), in press.
\bibitem{Poland} Poland, D.; Scheraga, H.A.
         Theory of Helix-Coil Transitions in Biopolymers.
         Academic Press, New York, 1970.
\bibitem{HO98c} Hansmann, U.H.E.; Okamoto, Y.
                Finite-Size Scaling of Helix-Coil Transitions in Poly-alanine
                Studied by Multicanonical Simulations. 
                J. Chem. Phys. 1999,  110, 1267-1276;  
                111,  1339(E).
\bibitem{AH99b} Alves, N.A.; Hansmann, U.H.E. 
                Partition Function Zeros and Finite Size Scaling
                  of Helix-Coil Transitions in a Polypeptide.
                Phys. Rev. Lett. 2000, 84, 1836-1839.
\bibitem{KHC99c} Kemp, J.P.;  Hansmann, U.H.E.;  Chen, Zh.Y.
                 Is There a Universality of the Helix-Coil Transition
                 in Protein Models ? 
                 Eur. Phy. J. B. 2000, 15, 371-374.
\bibitem{EC}  Sippl, M.J.;   N{\'e}methy, G.;   Scheraga, H.A. 
            Intermolecular Potentials from Crystal Data.
            6. Determination of Empirical Potentials for
            O-H$\cdots$O=C Hydrogen Bonds from Packing Configuration.
            J. Phys. Chem. 1984, 88, 6231-6233; and references therein.
\bibitem{MU} Berg, B.A.; Neuhaus, T. 
             Multicanonical Algorithms for First Order Phase Transitions.
             Phys. Lett. B. 1991, 267, 249-253.
\bibitem{AlvesIJMPC}  Alves, N.A.;  de Felicio, J.R.D.; Hansmann, U.H.E.
          A New Look at the 2d Ising Model from Exact Partition
          Function Zeros for Large Lattice Sizes.
          Int. J. Mod. Phys. C. 1997, 8, 1063-1072.
\bibitem{AlvesNPB92} Alves, N.A.; Berg, B.A.; Sanielevici, S.
         Spectral Density Study of the SU(3) Deconfining Phase Transition.
         Nucl. Phys. B. 1992, 376, 218-252.
\bibitem{Efron} Efron, B.
          Computers and the Theory of Statistics: Thinking the Unthinkable.
          SIAM Rev. 1979, 21, 460-480.

          Efron, B.; Tibshirani, R.J. 
          An Introduction to the Bootstrap. 
          Monographs on Statistics and Applied Probability.
          Chapman \& Hall.  1993, Vol. 57.
\bibitem{AGA} Alves, N.A.;  Hansmann, U.H.E.
              Helix Formation and Folding in an Artifical Peptide.
              Submitted for publication.
\bibitem{SMMP} Eisenmenger, F.;  Hansmann, U.H.E.;  Hayryan, Sh.; 
               Hu, C.K.
               [SMMP] A Modern Package for Simulation of Proteins.
               Comp.Phys. Comm. 2001, 138, 192-212.
\end{thebibliography}
\end{document}